\pdfoutput=1
\documentclass[%
    aps                ,   % Main styling  : [aps|aip]
    prl               ,   % Sub-styling   : [pra|prb|prc|prd|pre|prl|prstab|prstper|rmp] or [apl|bmf|cha|jap|jcp|jmp|rse|pof|pop|rsi]
    reprint            ,   % View style    : [preprint|reprint]
    twocolumn          ,   % Columns       : [onecolumn|twocolumn]
    showpacs           ,   % PACS numbers  : [showpacs|noshowpacs]
    superscriptaddress ,   % Author address: [superscriptaddress]
    nofootinbib        ,   % Footnotes     : [footinbib|nofootinbib]
    citeautoscript     ,   % Fix superscript citation order
    notitlepage        ,   % Don't take a whole page for the title
    floatfix
]{revtex4-1}
\usepackage[T1]{fontenc}
\usepackage[utf8]{inputenc}
\usepackage[english]{babel}
\setcitestyle{super}                 % Superscript citations (~ alt: "prb")
\addto\captionsenglish{\renewcommand{\figurename}{Fig.}}
\makeatletter
\renewcommand*{\fnum@figure}{{\normalfont\bfseries \figurename~\thefigure}}
\renewcommand*{\@caption@fignum@sep}{\textbf{ | }}
\makeatother

\usepackage{amsmath}
\usepackage{amssymb}
\usepackage{braket}
\usepackage{accents}
\usepackage{siunitx}
\usepackage[hyperref]{xcolor}
\usepackage{graphicx}
\usepackage[export]{adjustbox}
\graphicspath{{figures/PNG/}{figures/PDF/}{figures/}}

\usepackage[%hypertex,
    unicode           = true  ,
    plainpages        = false , 
    pdfpagelabels     = true  , 
    bookmarks         = true  ,
    bookmarksnumbered = true  ,
    bookmarksopen     = true  ,
    breaklinks        = true  ,
    backref           = false ,
    colorlinks        = true  ,
    linkcolor         = blue  ,
    urlcolor          = blue  ,
    citecolor         = red   ,
    anchorcolor       = green ,
    hyperindex        = true  ,
    linktocpage       = true  ,
    hyperfigures      = true
]{hyperref}
\hypersetup{
    linkcolor         = [RGB]{000 114 189} ,  % Strong Cornflower Blue
    urlcolor          = [RGB]{000 114 189} ,  % Strong Cornflower Blue
    filecolor         = [RGB]{000 114 189} ,  % Strong Cornflower Blue
    citecolor         = [RGB]{217 083 025} ,  % Brilliant Vermilion
    anchorcolor       = [RGB]{119 171 048} ,  % Moderate Spring Bud
    pdftitle={Multi-dimensional optical neural network},
    pdfauthor={Zhetao Jia <zhetao@berkeley.edu>}
}

\begin{document}

\title{Multi-dimensional optical neural network}

\author{Zhetao Jia}
\email{Corresponding author: zhetao@berkeley.edu}
\affiliation{Department of Electrical Engineering and Computer Sciences, University of California at Berkeley, Berkeley, California 94720, USA}

\author{Hector Rubio}
\affiliation{Electrical and Computer Engineering, Rochester Institute of Technology, Rochester, New York 14623, USA}

\author{Lilian Neim}
\affiliation{Microsystems Engineering, Rochester Institute of Technology, Rochester, New York 14623, USA}

\author{Jagang Park}
\affiliation{Department of Electrical Engineering and Computer Sciences, University of California at Berkeley, Berkeley, California 94720, USA}

\author{Stefan Preble}
\affiliation{Microsystems Engineering, Rochester Institute of Technology, Rochester, New York 14623, USA}

\author{Boubacar Kanté}
\email{Corresponding author: bkante@berkeley.edu}
\affiliation{Department of Electrical Engineering and Computer Sciences, University of California at Berkeley, Berkeley, California 94720, USA}
\affiliation{Materials Sciences Division, Lawrence Berkeley National Laboratory, Berkeley, California 94720, USA}

\date{\today}


\begin{abstract}
The development of deep neural networks is witnessing fast growth in network size, which requires novel hardware computing platforms with large bandwidth and low energy consumption. Optical computing has been a potential candidate for next-generation computing systems. Specifically, wavelength-division multiplexing (WDM) has been widely adopted in optical neural network architecture to increase the computation bandwidth. Although existing WDM neural networks architectures have shown promise, they face challenges in the integration of light sources and further increase of the computing bandwidth. To overcome these issues, we introduce a mode-division multiplexing (MDM) strategy, offering a new degree of freedom in optical computing within the micro-ring resonator platform. We propose a MDM approach for small-scale networks and a multi-dimensional architecture for large-scale applications, supplementing WDM with MDM to enhance channel capacity for computations. In this work, we design and experimentally demonstrate key components for the MDM computing system, i.e., a multimode beam splitter, a thermo-optical tuner for the high-order mode, and a multimode waveguide bend. We further show a 2-by-2 matrix multiplexing system fabricated in a foundry that works for both MDM and MDM-WDM computing, which confirms that our approach successfully increases the input vector size for computing and ensures compatibility with existing WDM networks.
\end{abstract}

\maketitle

%%%%%%%%%%%%%%%%%%%%%%%%%%  body  %%%%%%%%%%%%%%%%%%%%%%%%%%
\section{Introduction}

As conventional electronic computing platforms approach performance limits, photonic-based computing emerges as a promising alternative, offering potential reductions in energy consumption, enhanced bandwidth, and decreased latency. Therefore, there has been a lot of effort and interest in developing the next generation of computing platforms based on photonics \cite{Huang2021a, Xu2021, Feldmann2021a, VanNiekerk2022b, Wright2022, Zhou2022, Pai2023, Rizzo2023, Xu2022, Filipovich2022, Wang2022, Luan2023, Wang2023}. A critical concern for any novel computing platform is scalability, encompassing factors like power consumption, operational speed, and signal-to-noise ratio. Several integrated photonics platforms have been proposed for scalable optical neural networks, including the coherent nanophotonic circuit based on Mach-Zender interferometers \cite{Shen2017}, and the wavelength-division multiplexing (WDM)-based neural network using micro-ring resonators \cite{Tait2017}. A recent study has systematically analyzed the scaling law for power consumption depending on the system size and bandwidth \cite{Tait2022}. While the WDM approach with micro-ring resonators has the potential to enable low-latency optical computing with a smaller system footprint, the integration of multi-wavelength laser sources can be challenging in material integration and wavelength stabilization. Despite the recent development of the on-chip frequency comb as a possible solution for the laser source \cite{Gaeta2019, Chang2022, Shu2022, Bai2023, Rizzo2023}, it has been shown that the packing wavelength crossing below $100GHz$ can lead to significant crosstalk\cite{Padmaraju2014}, which puts an upper limit on the number of channels. Recently, 32 channels WDM have been implemented \cite{Rizzo2023}. However, the number of available channels in existing photonic computing platform is still insufficient for modern neural network architectures. For example, a simple feed-forward neural network for the MNIST dataset (28 by 28 pixels) has an input layer of 784 units \cite{Deng2012}, while a kernel for the convolution neural network can takes up to 27 channels (3 by 3 pixels and 3 colors). To address the bandwidth bottleneck, an approach based on MDM-WDM has been demonstrated to be a promising way to increase the bandwidth of optical communication\cite{Luo2014a, Wu2017, Yang2022}. The question is: can we adopt mode multiplexing for optical computing\cite{Gordon2018}? We propose and demonstrate the integration of mode-division multiplexing (MDM) as a feasible new dimension for enhancing optical computing capabilities. We report a mode-division multiplexing optical matrix multiplication system fabricated on an AIM Photonics Multi-Project Wafer (MPW) \cite{Fahrenkopf2019} and validate the signal and crosstalk. Our experimental results show that MDM can increase the input vector size, and, more importantly, it is compatible with existing wavelength-division multiplexed neural networks for increasing the dimension of matrix multiplication.

\section{Design architecture}

\begin{figure*}[h!]
    \centering
    \includegraphics[width=0.9\textwidth]{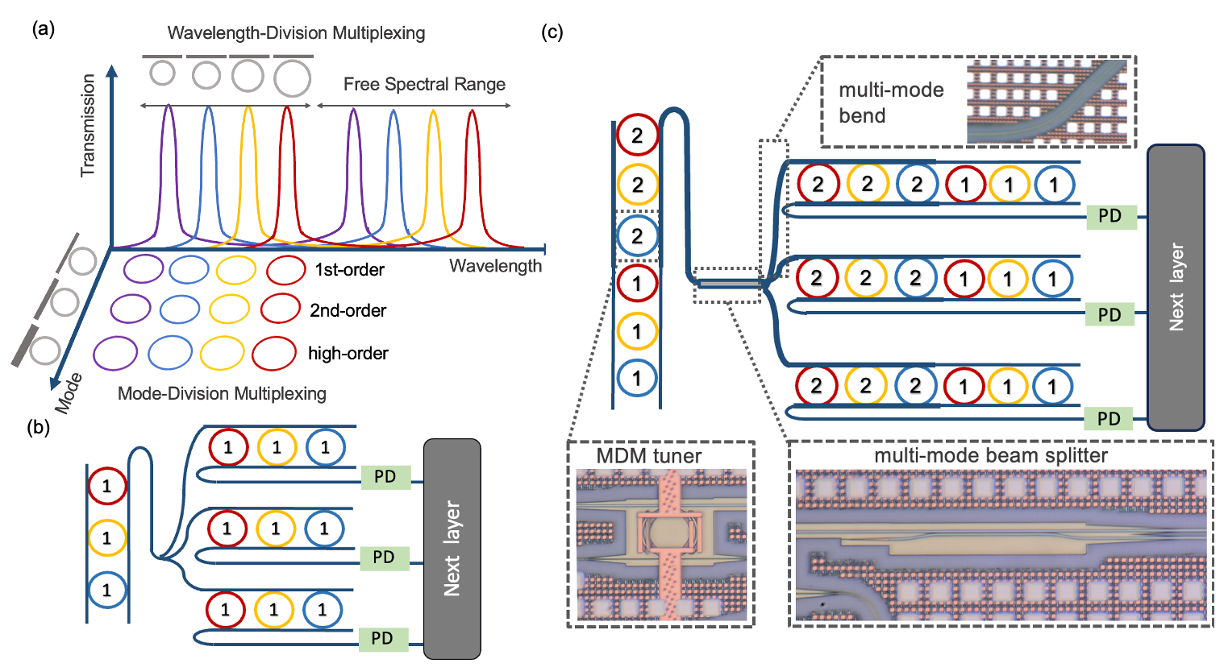}
    \caption{Architecture of MDM-WDM optical neural network. (a) Schematics of multi-dimensional optical computing using both wavelength-division multiplexing and mode-division multiplexing. The matrix multiplication is limited by the number of wavelengths within one free-spectral range in a conventional WDM-based optical neural network system. (b) WDM-based optical neural network and (c) WDM-MDM-based optical neural network. Inset: microscope images of the MDM thermo-optical tuner, multimode beam splitter, and low-loss multimode waveguide bend fabricated by AIM Photonics.}
    \label{fig:Fig1_schematics}
\end{figure*}

We propose a multi-dimensional optical neural network architecture as shown in Fig.~\ref{fig:Fig1_schematics}. In traditional optical neural networks utilizing WDM, minimum wavelength spacing and the wavelength range determine the total number of independent channels for computing. In Fig.~\ref{fig:Fig1_schematics}(a), MDM introduces an additional degree of freedom, alongside WDM, to create more orthogonal channels, thereby expanding the capacity for optical computing. Our MDM approach uses micro-ring-based thermo-opical tuners that each couples a high-order mode from the bus waveguide, as represented by concentric rings in the schematics. While the conventional neural network allows an optical matrix-vector multiplication for an input vector of dimension $N$, which is encoded in different wavelengths shown in Fig.~\ref{fig:Fig1_schematics}(b), MDM-WDM-based neural network enables a much larger matrix operator of dimension $MN$, where $M$ is the number of modes available. The schematics in Fig.~\ref{fig:Fig1_schematics}(b) illustrates a case with $N=3$ and $M=2$. Implementing this higher-dimensional optical computing system necessitates the design and optimization of three crucial components: a thermo-optical tuner for the high-order mode, a multimode beam splitter, and a multimode waveguide bend (Fig.~\ref{fig:Fig1_schematics}(c)). The multimode beam splitter is needed to distribute the intensity of each mode equally into the row of the multiplication matrix, and the multimode waveguide is crucial for minimizing the scattering loss, especially for high-order modes. The microscope images for corresponding fabricated components are shown in the zoomed-in views of Fig.~\ref{fig:Fig1_schematics}(c).

\section{Component-level: simulation and experimental results}

To demonstrate the proposed mode-division multiplexed neural network, we designed and simulated the three key components mentioned above. In this section, we elaborate on the simulation and experimental results of the MDM thermo-optical tuner for a high-order mode. The results for the multimode beam splitter and multimode waveguide bend are presented in supplementary materials. 

\begin{figure*}[htbp]
    \centering
    \includegraphics[width=0.9\textwidth]{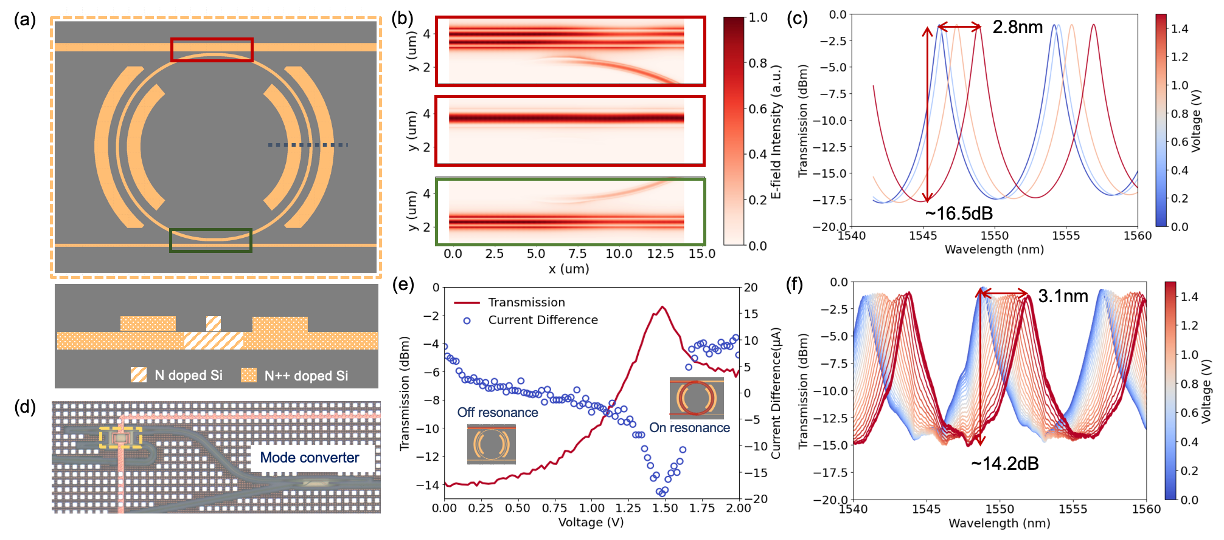}
    \caption{MDM thermo-optical tuner simulation and measurement results. (a) The top view of the MDM tuner with the asymmetric coupler (red box) for the $TE_{01}$-mode, and the symmetric coupler (green box) for the fundamental mode. The cross-section view on each side of the micro-ring is shown. (b) Simulated intensity profiles show that the $TE_{01}$-mode couples into the micro-ring from the asymmetric coupler, while the fundamental mode ($TE_{00}$) goes through with negligible coupling. The symmetric coupler has a similar coupling coefficient. (c,f) Simulation and experimental results for the thermal shift of the resonances of the MDM tuner. (e) Photo-conductive effect for calibration of the MDM tuner. The peak in transmission matches the maximum current difference between the laser on and off. 
    }
    \label{Fig2_multimode_tuner}
\end{figure*}

The designed MDM thermo-optical tuner controls the coupling and conversion from the quasi-$TE_{01}$ mode to the quasi-$TE_{00}$ mode of the input waveguide. We will refer to them later as $TE_{01}$ mode and $TE_{00}$ mode, respectively, for simplicity. As illustrated in Fig.~\ref{Fig2_multimode_tuner}(a), the designed thermo-optical tuner consists of ridge-shaped waveguides to allow the doped section to conduct current for both thermal tuning and calibration purposes. Specifically, we utilize the photoconductive effect \cite{Jayatilleka2015a} to monitor the state of the tuners for the high-order mode, which will be explained in more details later in this section. In the coupling region, the tuner consists of an asymmetric coupling region ($TE_{01}$ to $TE_{00}$) at the top (red box) and a symmetric coupling region ($TE_{00}$ to $TE_{00}$) at the bottom (green box). The width of the input multimode coupler is designed to match the effective index of the high-order mode ($TE_{01}$) with the fundamental mode inside the micro-ring to selectively couple the $TE_{01}$ mode from the input. We simulate multimode waveguide of width 1.1 $\mu m$, obtaining a coupling efficiency of $24.2\%$ near $1550nm$ wavelength with a coupling length of 1 $\mu m$ for the straight section. The field profile with $TE_{01}$ mode input is shown in Fig.~\ref{Fig2_multimode_tuner}(b). In contrast, we show that only the $TE_{01}$ mode is coupled but the fundamental mode propagates through the junction with negligible crosstalk. The symmetric coupler at the bottom (green box) couples the fundamental mode inside the micro-ring tuner to the single-mode waveguide with a similar coupling efficiency of $23.5\%$ (Fig.~\ref{Fig2_multimode_tuner}(b), bottom). We design the two coupling coefficients to be similar for a nearly critically coupled condition such that the input intensity can be fully coupled to the drop port when the device is on resonance. The circuit simulation of the thermo-optical tuner shows that an applied voltage of 1.5 $V$ leads to a wavelength shift of 2.8 $nm$, which can modulate the transmission by ~16.5 $dB$ (Fig.~\ref{Fig2_multimode_tuner} (c)). We experimentally validate the simulated MDM tuner with a testing device consisting of the mode converter followed by the tuner as shown in Fig.~\ref{Fig2_multimode_tuner}(d). The mode converter transforms the fundamental mode from the input to $TE_{01}$ at the output, which is sent to the tuner to couple back to the fundamental mode that can be directly probed through the edge coupler. Experimental results (Fig.~\ref{Fig2_multimode_tuner}(f)) show an extinction ratio of $-14.2dB$, with a resonance shift of $3.1\ nm$ at a voltage difference of $1.5V$, which is consistent with the simulation result.

One of the major challenges of scalable weight banks with an MDM thermo-optical tuner array is the calibration of the resonance wavelengths. Such calibration is trivial for a single heater-based resonator if the optical output can be directly measured in a feedback loop. However, as the dimension of weight matrices and the number of layers in the neural network increase, the calibration of each tuner with optical response becomes challenging. A built-in current sensor for calibration based on photoconductive effect was proposed \cite{Jayatilleka2015a} and then adopted for weight banks in optical neural networks \cite{Tait2018b, Huang2020, Huang2021}. We introduce the photoconductive tuner to couple a high-order mode as shown in the cross-section in Fig.~\ref{Fig2_multimode_tuner}(a), where the regions on both sides of the micro-ring are N-doped and the central region of the micro-ring is lightly doped.  
% AIM \cite{Fahrenkopf2019}. 
% why the current change is negative??
In Fig.~\ref{Fig2_multimode_tuner}(e), the transmission at the drop port is maximized near applied voltage $1.5V$, where the micro-ring resonator is on resonance. The coupling of light can be directly probed in the current change by switching the laser on and off. When the laser is on and the micro-ring is on resonance, the photoconductive effect leads to a change in conductance, compared to the case where the laser is off.  The alignment between the maximum current difference and the highest transmission at the drop port (see supplementary information) indicates that the calibration of transmission can be achieved without measuring the optical output. Therefore, the system would have a built-in calibration mechanism to compensate for the fabrication imperfection across different tuners. It is worth noting that contrary to previous works \cite{Jayatilleka2015a, Tait2018b, Huang2020, Huang2021}, we observed a negative change in conductance when the micro-ring tuner is on resonance, which can be caused by the thermal effect from the higher doping concentration ($10^{18} cm^{-3}$) used.

\section{System-level: simulation and experimental results}

\begin{figure*}[htbp]
    \centering
    \includegraphics[width=0.9\textwidth]{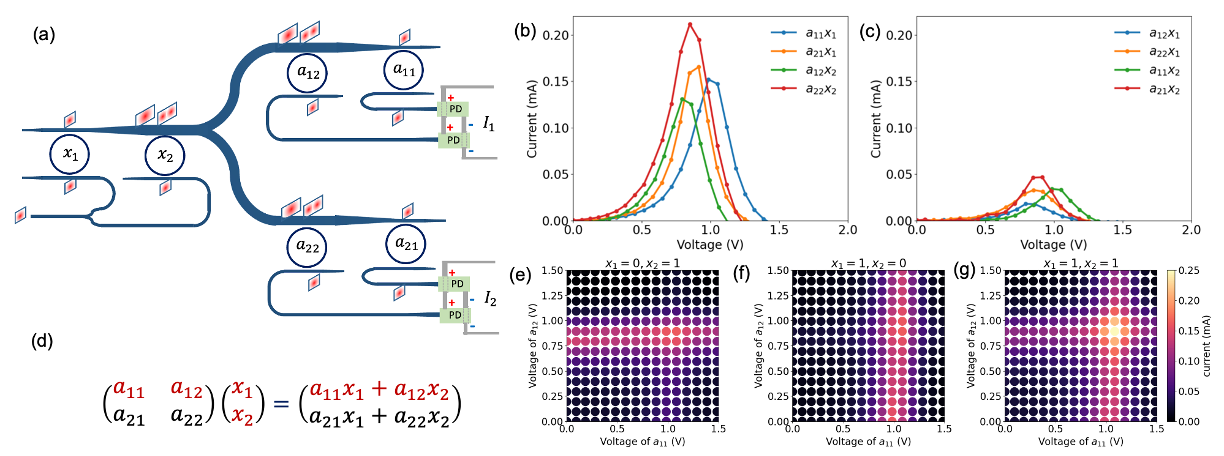}
    \caption{Matrix multiplication with MDM thermo-optical tuners. (a) Schematic figure for the implemented 2 by 2 matrix multiplication with MDM. The fabricated structure consists of six micro-rings, two for controlling the amplitudes of inputs and four for matrix multiplication. Two photodetectors are connected in parallel for summing up the measured current for each row. (b) Measured photodetector currents for different combinations of input and weight micro-rings, showing the results for element-wise multiplication. (c) Measured cross-talk current for different combinations. (d-f) Multiplication and addition results for different input combinations (1,0), (0,1), and (1,1).
    }
    \label{Fig3_experiment_MDM}
\end{figure*}

In this section, we demonstrate a 2 by 2 matrix multiplication unit using the MDM scheme. The schematic figure of the fabricated system is shown in Fig.~\ref{Fig3_experiment_MDM}(a). The laser input is coupled to the bus waveguide as $TE_{00}$ and $TE_{01}$ mode with input tuner $x_1$ and $x_2$, respectively. The bus waveguide is connected to the multimode beam splitter, and then to four micro-rings for a matrix multiplication operation. The micro-rings convert the target mode back into the fundamental mode and are measured by the photodetectors. We first show the measurement results for element-wise multiplication  (i.e.  $a_{11} x_1, a_{12} x_2, a_{21} x_1, a_{22} x_2$). For simplicity, the input tuners are tuned to couple maximum intensity into the bus waveguide in each case. The current measured by the on-chip photo-detector is monitored while modulating the weight rings in the matrix. In Fig.~\ref{Fig3_experiment_MDM}(b), we demonstrate that the target modes can be coupled into the photodetectors as expected. Crosstalk between different modes is visualized in Fig.~\ref{Fig3_experiment_MDM}(c), where we estimate an average signal-to-noise ratio of 5, based on the measurement results. The mismatch between the simulation and the fabricated device can introduce unwanted intermodal coupling at the micro-ring tuner and the multi-mode directional coupler, which can be reduced by optimizing the width of the multimode waveguide and the coupling length of the multimode directional coupler in future iterations. The multiplication and addition results are shown in Fig.~\ref{Fig3_experiment_MDM}(e-g) with different combinations of the input. For an input vector of $(0,1)$ where only the $TE_{01}$ mode is coupled into the bus waveguide and distributed through the splitter, the current is determined by the applied voltage on the corresponding tuner $a_{12}$, which is designed to couple the $TE_{01}$ mode. Similarly, the current response is controlled by the voltage applied on tuner $a_{11}$ with an input vector $(1,0)$. For the $(1,1)$ case, the current is summed up as expected, demonstrating a multiplication-and-addition operation.

\begin{figure*}[htbp]
    \centering
    \includegraphics[width=0.9\textwidth]{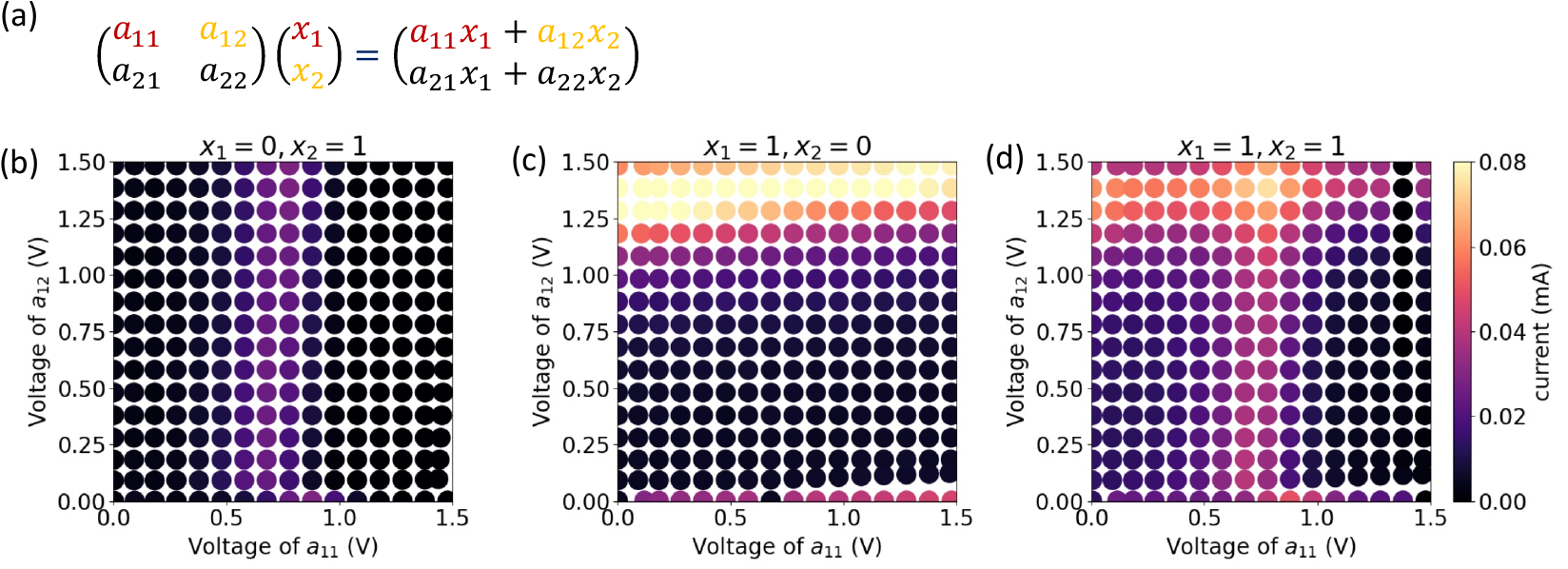}
    \caption{Matrix multiplication with MDM-WDM scheme. (a) Multiplication and addition results for different input combinations (1,0), where only the laser 1 is on. (b) Multiplication and addition results with laser 2 on. (c) Multiplication and addition results with both lasers on.
    }
    \label{Fig4_experiment_WDMMDM}
\end{figure*}

We show that the proposed MDM scheme is compatible with WDM. We combine two lasers and send them to the input channel. In Fig.~\ref{Fig4_experiment_WDMMDM}, the output current is shown for different input combinations. When laser 1 is on and laser 2 is off, the fundamental mode at $\lambda_1 = 1472.6nm$ is coupled, leading to the output current modulated by tuner $a_{11}$. On the other hand, when laser 2 is switched on, the input laser of wavelength $\lambda_2 = 1550.7nm$ is coupled to the second order mode of the bus waveguide and therefore modulated by the $a_{12}$. In Fig.~\ref{Fig4_experiment_WDMMDM}(c), we show that the output current at different wavelengths and modes can be summed up, resulting in the addition of the current. These results confirm that MDM and WDM can be effectively combined for computation, and can potentially greatly increase the number of channels available for matrix multiplication purposes.

\section{Conclusion}

We proposed and demonstrated a mode-division multiplexing optical matrix multiplication platform. We designed and characterized the necessary components following the foundry fabrication constraints, including a photoconductive thermo-optical tuner for the high-order mode, a multimode beam splitter, and a multimode waveguide bend, for a scalable WDM-compatible neural network. We experimentally characterized a matrix multiplication, and demonstrated the element-wise multiplication and row addition. The crosstalk between different modes is characterized and can be further reduced by optimizing the width of the multimode waveguide and the coupling length of the multimode beam splitter. Such mode-division architecture can be used for small-size neural networks like convolutional neural networks for the prepossessing of graphs, which can reduce the complexity of multiple laser/frequency comb integration. For large-scale neural network applications, we showed that the proposed mode-division multiplexing is compatible with wavelength division multiplexing to increase the dimension of channels as well as the matrix. Conclusively, the proposed platform shows promise for developing compact, low-energy, and scalable computing systems, with the potential for future optimization and application in large-scale neural networks.

% \begin{backmatter}
\section{Acknowledgments}

This material is based on research sponsored by the Air Force Research Laboratory under AIM Photonics (agreement number FA8650-21-2-1000).  The U.S. Government is authorized to reproduce and distribute reprints for Governmental purposes notwithstanding any copyright notation thereon. The views and conclusions contained herein are those of the authors and should not be interpreted as necessarily representing the official policies or endorsements, either expressed or implied, of the United States Air Force, the Air Force Research Laboratory or the U.S. Government.

B.K. acknowledges support from the Army Research Office (ARO) (W911NF2310027); the Moore Inventor Fellows programme; the Bakar Fellowship at UC Berkeley, and the National Science Foundation (OMA-2016245, 2137645).

% \section{Disclosures}
% The authors declare no conflicts of interest.

% \section{Data availability} Data underlying the results presented in this paper are not publicly available at this time but may be obtained from the authors upon reasonable request.

% \bmsection{Supplemental document}
% See Supplement 1 for supporting content. 

% \end{backmatter}

%For small-scale optical neural networks, it is possible to replace WDM fully with the MDM method to eliminate the difficulty of multi-wavelength source integration.

%%%%%%%%%% If using BibTeX:
\bibliography{references}

% Include pre-compiled Supplementary Material
% \includesupplementary
\end{document}